\documentclass[sigconf,10pt,nonacm]{acmart}

\usepackage{graphicx}
\usepackage{endnotes}
\usepackage{xspace}
\usepackage{epigraph}
\usepackage{enumitem}
\usepackage{soul}  

\newcommand{\system}{\textsc{Guillotine}\xspace}

\usepackage{hyphenat} 
\usepackage{xcolor} 
\usepackage{url}

\copyrightyear{2025}
\acmYear{2025}
\setcopyright{rightsretained}
\acmConference[HOTOS '25]{Workshop on Hot Topics in Operating Systems}{May 14--16, 2025}{Banff, AB, Canada}
\acmBooktitle{Workshop on Hot Topics in Operating Systems (HOTOS '25), May 14--16, 2025, Banff, AB, Canada}
\acmDOI{10.1145/3713082.3730391}
\acmISBN{979-8-4007-1475-7/2025/05}

\begin{document}

\settopmatter{printfolios=true}

\date{}

\title{\system: Hypervisors for Isolating Malicious AIs}
\titlenote{Accepted to the ACM SIGOPS 2025 Workshop on Hot Topics in Operating Systems.}

\author{James Mickens}
\email{mickens@g.harvard.edu}
\affiliation{%
  \institution{Harvard University}
  \country{}
}

\author{Sarah Radway}
\email{sradway@g.harvard.edu}
\affiliation{%
  \institution{Harvard University}
  \country{}
}

\author{Ravi Netravali}
\email{rnetravali@cs.princeton.edu}
\affiliation{%
  \institution{Princeton University}
  \country{}
}

\newcommand{\longpage}{\enlargethispage{\baselineskip}}
\newcommand{\shortpage}{\enlargethispage{-\baselineskip}}

\maketitle


\pagenumbering{gobble}

\subsection*{Abstract}
As AI models become more embedded
in critical sectors like
finance, healthcare, and the military,
their inscrutable behavior
poses ever-greater risks to society.
To mitigate this risk,
we propose Guillotine,
a hypervisor architecture for sandboxing
powerful AI models—models that,
by accident or malice,
can generate existential threats to humanity.
Although Guillotine borrows
some well-known virtualization techniques,
Guillotine must also introduce
fundamentally new isolation mechanisms
to handle the unique threat model
posed by existential-risk AIs.
For example, a rogue AI may try to
introspect upon hypervisor software
or the underlying hardware substrate
to enable later subversion of that control plane;
thus,
a Guillotine hypervisor requires careful co-design
of the hypervisor software and
the CPUs, RAM, NIC, and storage devices
that support the hypervisor software,
to thwart side channel leakage
and more generally eliminate mechanisms for AI to exploit reflection-based vulnerabilities.
Beyond such isolation at
the software, network, and microarchitectural layers,
a Guillotine hypervisor must also provide
physical fail-safes more commonly associated with
nuclear power plants, avionic platforms,
and other types of mission-critical systems.
Physical fail-safes, e.g.,
involving electromechanical disconnection of network cables,
or the flooding of a datacenter which holds a rogue AI,
provide defense in depth if
software, network, and microarchitectural isolation
is compromised and a rogue AI
must be temporarily shut down or permanently destroyed.

\section{Motivation}
\label{sec:intro}

A machine learning model
tries to emulate human reasoning.
To do so,
a model encodes observations about training data
using numerical parameters and
links between those parameters.
Current state-of-the-art models
are so large that
their internal organization
is opaque to their human creators.
For example,
the open-source BLOOM model
has 176 billion parameters~\cite{bloom},
the open-source Llama 3.1 model
has 405 billion parameters~\cite{llama3-1},
and the closed-source GPT-4 model
is rumored to have
more than a trillion parameters~\cite{chatgpt-params}.

\shortpage

Humans cannot directly understand
the relationships between
such vast constellations of parameters.
Automated methods for understanding those relationships
(and how they generate model outputs)
are an active area of research.
Unfortunately,
such model interpretability techniques
appear to be inherently fragile.
Consider the task of explaining LLM inferences.
The soundness of LLM interpretability techniques
is vulnerable to
instabilities in the underlying LLM itself~\cite{singh2024}.
For example,
the fact that LLMs are sensitive
to minor changes in prompt phrasing~\cite{salinas2024}
can result in
a model's self-reported chain-of-thought
being an unfaithful representation
of the model's actual reasoning process~\cite{turpin2023}.
Furthermore,
an LLM's tendency to hallucinate~\cite{das2024}
can manifest itself not only in
the model's answer to a question,
but in the model's
explanation for that answer~\cite{durrett2022}.

The opacity of model reasoning
is troubling because
models are increasingly connected to
societally important infrastructure.
For example,
in financial settings,
misbehaving models
can generate huge monetary losses
due to bugs.
Those bugs might have been
unintentionally introduced
by model makers~\cite{min2022}
or intentionally induced
by adversarial examples~\cite{xie2022}.
In warfighting scenarios,
military leaders
are already concerned
that AI-governed weapons
may escalate conflicts
due to ignorance of
geopolitical nuance~\cite{rand-military-ai};
these escalation problems
are exacerbated when AI makes decisions
too quickly for humans to
review those decisions~\cite{lin-greenberg-2020}.

\longpage
\longpage

Model alignment techniques~\cite{wolf2024}
try to ensure that
models adhere to human-defined behavioral norms.
However (and concerningly),
models can fake alignment compliance during training
to later act in
non-aligned ways post-deployment~\cite{greenblatt2024}.
Researchers have also demonstrated that,
in the specific context of LLMs,
if model alignment
does not completely eliminate
the possibility of an undesirable model behavior,
an adversarial prompt
can always elicit that behavior~\cite{wolf2024}.
Thus,
society faces an increasing risk
that an artifical general intelligence (AGI) model
which matches or exceeds human reasoning
will generate catastrophic harms
in real life.

\shortpage

In this paper,
we argue that the systems community
has a critical role
in preventing AGI-based existential harms;
alignment techniques at training time
must be supplemented by
sandboxing mechanisms at inference time.
We propose a new class of hypervisor,
called a \system hypervisor, 
for isolating AGI inference code.
Due to the unique threats posed by rogue models,
and the potentially catastrophic risks
of not containing those threats,
\system's design
looks much different than
that of a traditional hypervisor.
We hope that this paper
encourages the system community
to participate more actively
in discussions about existential AGI risks.

\section{High-level \system Design}
\label{sec:design-high-level}

{\bf Background:} A \textbf{model service}
is a distributed system
that accepts inference requests
and outputs inference results.
Internally,
the service has 
one or more request queues,
and one or more replicas
of each model
whose computations will contribute
to an inference~\cite{clipper,clockwork,nexus,shepherd}.
Computations are split between
CPUs and GPUs,
with GPUs typically doing the bulk of the inference work.
CPUs load-balance requests
across different GPUs,
and orchestrate the transfer of requests and responses
between CPU DRAM and on-GPU DRAM.
CPUs also manage various caches,
e.g., LLMs key/value caches,
located in GPU DRAM,
which store previously-generated tokens
as well as intermediate values
output by different LLM layers~\cite{vLLM, cachedAttention,marconi}.
CPUs also manage a model's access
to external services and IO devices.
For example,
LLMs often perform
retrieval-augmented generation,
supplementing a user-supplied prompt
with information from
a database of domain-specific document embeddings~\cite{ragCache,ragServe}.


\begin{figure}[t!]
\centering
\includegraphics[width=\columnwidth]{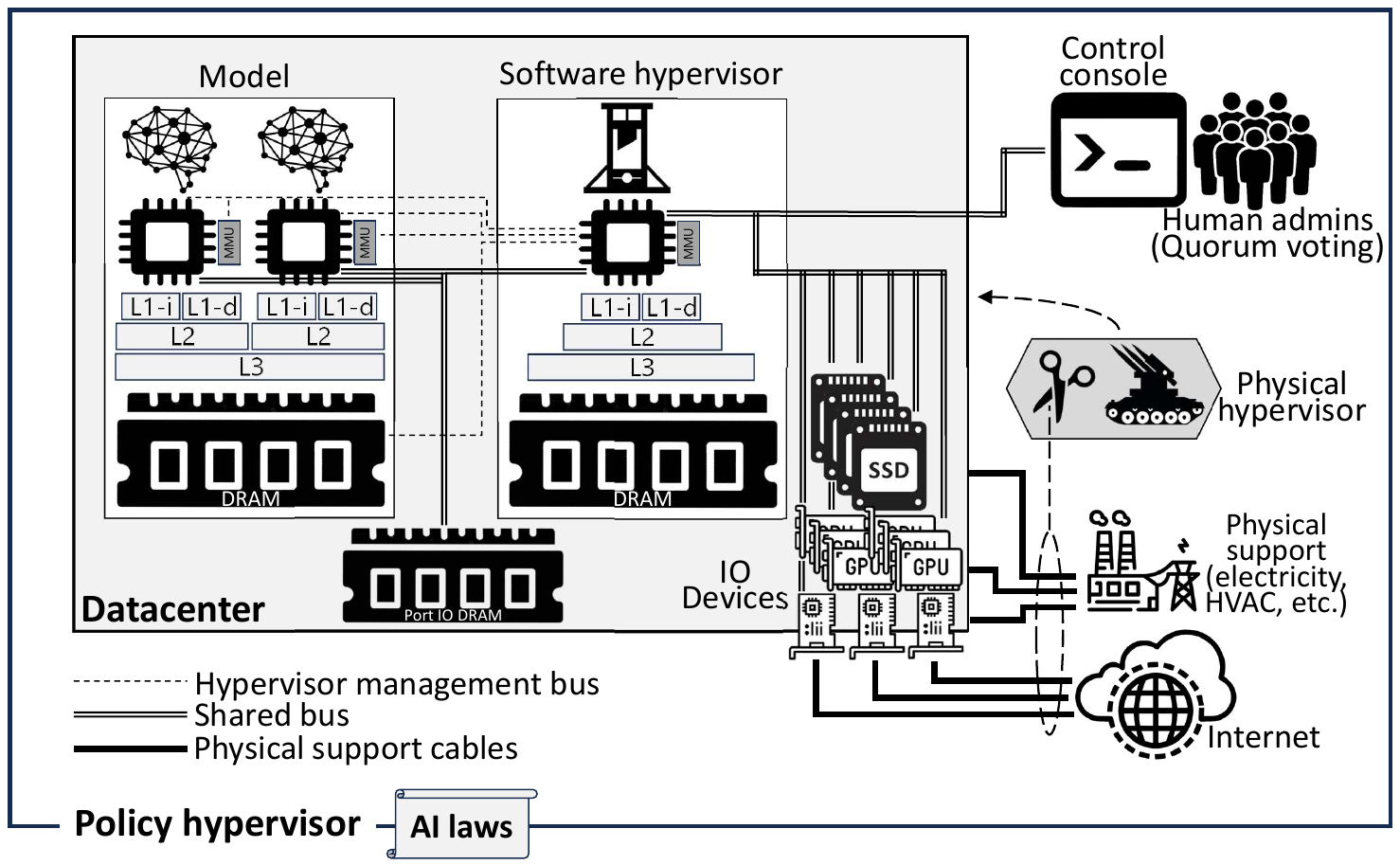}
\vspace{-7mm}
\caption{The \system sandboxing architecture.}
\vspace{-5mm}
\label{fig:high-level-design}
\end{figure}

$ $\\
\noindent
{\bf \system:}
Given such a model service,
Figure~\ref{fig:high-level-design} depicts
the high-level design
of a \system hypervisor.
\begin{itemize}[leftmargin=*]
  \item The \textbf{model}
        interacts with the outside world
        via the \textbf{guest API}
        of the \textbf{\system software hypervisor}
        (\S\ref{sec:hypervisor-sw}).
  \item The software hypervisor executes atop
        the \textbf{\system microarchitectural hypervisor}.
        (\S\ref{sec:hypervisor-uarch}).
        Conceptually,
        the microarchitectural hypervisor
        is the \system analog to
        traditional hardware-level virtualization mechanisms
        like Intel VT-x~\cite{uhlig2005}.
        However,
        \system's microarchitectural hypervisor
        requires qualitatively different in-silicon support
        to achieve \system's desired isolation properties.
        For example,
        to restrict the ability of a model
        to infer the computational state
        of the software-level hypervisor,
        \system requires extremely limited
        microarchitectural co-tenancy
        between the model and
        the software-level hypervisor;
        this approach eliminates many kinds of
        side-channel leakages~\cite{vicarte2021}
        by definition.
  \item The \textbf{\system physical hypervisor}
        (\S\ref{sec:hypervisor-phys})
        implements physical fail-safes
        more commonly associated with
        rockets,
        nuclear reactors,
        and other types of mission-critical systems.
        These fail-safes,
        e.g., involving electromechanical disconnection
        of a datacenter's network cables,
        or the flooding of a datacenter
        which holds a rogue AI,
        provide defense in depth
        if \system's microarchitectural and software-level
        isolation is compromised
        and a rogue AI must be
        temporarily shut down or permanently destroyed.
  \item A \system hypervisor is only effective
        if AI models are forced to run atop \system.
        Thus,
        the last part of the \system sandbox
        is a \textbf{policy hypervisor}---a set of
        legal regulations which
        (1) provide formal specifications
        for how \system-class hypervisors
        must be built, and
        (2) require potentially dangerous models
        to run atop \system infrastructure.
\end{itemize}



\section{\system Design}
\label{sec:design}


\subsection{Threat model}
\label{sec:hypervisor-threat-model}


In our threat model,
the attacker is
a \textbf{deployed, post-training AI model}
which receives \textbf{inputs},
computes on those inputs,
and then generates \textbf{outputs}.
\system's goal
is to prevent the model
from escaping \system's sandbox
by accident or intentional malice.
A critical attack vector to thwart
is an actively malicious model
being able to introspect upon the \system sandbox
to learn how to escape.
Another important attack to prevent
is a recursively self-improving model
which seeks a runaway gain in intelligence.

Some model inputs may be submitted
by external-to-the-model entities,
e.g., a human or another computer
that has sent an inference request.
Other inputs
may be fetched by the model itself.
For example,
as the model ponders a query,
the model may issue
a database read
to fetch query-specific contextual information~\cite{ragSurvery2024}.
A model may send outputs to
networks, storage devices,
computational accelerators (e.g., GPUs),
or physical actuators
(e.g., when models control industrial equipment).
\system uses the abstract term ``\textbf{port}''
to refer to a channel that
delivers inputs to or send outputs from
a model.
%
\system requires the ability to
monitor and control all ports
that are visible to a model.

\system's \textbf{trusted computing base}
consists of
\system's software-level hypervisor (\S\ref{sec:hypervisor-sw})
and the \system silicon
which implements
ISA-level and microarchitectural\hyp level
isolation mechanisms (\S\ref{sec:hypervisor-uarch}).
\system's physical safeguards (\S\ref{sec:hypervisor-phys})
are also in the trusted computing base.
\system's \textbf{human administrators}
are partially trusted;
as we describe in Section~\ref{sec:hypervisor-phys},
important management decisions in \system
require voting among a set of admins.

Defining ``rogue AI behavior''
is a subtle matter involving interdisciplinary reasoning.
In this paper,
we focus on
providing a multi-level hypervisor
that can do two things reliably:
(1) provide system-level observations
about a model's behavior
to a black-box (from \system's perspective)
detector for malicious AI activity;
(2) receive a verdict from that detector
(or human administrators),
and then change
the model's isolation level
as appropriate.
The \textbf{black-box misbehavior detector}
is in the \system TCB;
we sketch possible detectors
in Section~\ref{sec:hypervisor-sw}.
\subsection{Microarchitectural Hypervisor}
\label{sec:hypervisor-uarch}


A traditional virtualization-aware processor~\cite{uhlig2005,sa2023,dall2018}
runs both guest code and hypervisor code.
At the microarchitectural level,
hardware structures like
functional units, branch predictors,
TLBs, and memory caches
may contain state
from both the guest and the hypervisor~\cite{smt},
even though,
at the ISA level,
privilege modes restrict
the ISA-visible state that
is accessible to guest code.
The mingling of state
enables side channels that
can leak hypervisor state to the guest~\cite{foreshadow-ng,spectre}.

Those side channels are unacceptable to \system,
because a key goal of \system
is to prevent a malicious model
from discovering
the implementation details of
the \system isolation environment
(\S\ref{sec:hypervisor-threat-model}).
\system also requires a pervasive ability
to inspect all aspects
of a model's computational state.
Thus,
\system requires special microarchitectural support.

First,
with respect to the physical CPU cores
on a single machine,
\system distinguishes between
\textbf{hypervisor cores} and
\textbf{model cores}.
Software-level hypervisor code only runs on
hypervisor cores;
post-initialization,
a model core only runs model code.
Model cores are connected via buses
to a memory hierarchy
containing DRAM and L1--L3 caches.
Hypervisor cores
have their own memory hierarchy
that is inaccessible
to model core buses.
A hypervisor core
also has dedicated, model-inaccessible control buses
which allows the hypervisor core
to exert management privileges
over model cores.
In particular,
a hypervisor core can:
\vspace{-1mm}
\begin{itemize}[leftmargin=*]
  \item forcibly pause
        all operations on a model core;
  \item inspect and modify
        the ISA-level
        state of a halted 
        core;
  \item set watchpoints on model code or memory locations;
  \item configure the model core's MMU
        such that the model cannot
        create new executable pages
        or write to old executable pages;\footnote{This approach
        is similar to Apple's mechanisms for
        kernel integrity protection and
        system coprocessor integrity protection~\cite{kipScip}.
        Implementing such protection
        is cheap at the hardware level---essentially,
        the MMU just tracks base+bound information
        for valid executable regions,
        and disallows PTE configurations that would
        enable read access to those regions
        or create new executable pages
        outside of those regions.
        Hypervisor cores also
        lock their MMUs in this fashion
        after loading the hypervisor software.}
        this prevents the model from
        injecting new code at runtime
        for the purpose of recursive self-improvement;
  \item forcibly clear all microarchitectural state
        on that core
        (to delete information that
        a model might try to hide via
        microarchitectural\hyp level
        covert channels~\cite{wistoff21})\footnote{In the
        covert channel,
        the model would be
        both the sender and the receiver!};
  \item single-step or fully resume a halted model core; and
  \item forcibly power-down a halted model core.
\end{itemize}
A hypervisor core
also has a private bus,
connected to model-core memory,
to enable reading and writing
the DRAM of halted model cores.
%

Note that
the \system approach
simplifies some aspects of processor design.
For example:
\begin{itemize}[leftmargin=*]
  \item A model core does not require
        extended page table (EPT) support~\cite{bhargava2008}.
        On a traditional non-\system core,
        one or more guests
        share the same DRAM that is also used
        by the hypervisor;
        EPTs allow the hypervisor to control
        which portions of physical DRAM
        are shared with each guest.
        However,
        in \system,
        a model core
        lacks the physical buses
        needed to access hypervisor DRAM,
        so EPTs are unnecessary to enforce memory isolation.
  \item A \system model core
        can also avoid the complexity
        of virtualizing interrupts and exceptions.
        Software on the model core
        handles locally-generated interrupts and exceptions
        (e.g., the timer interrupt and
        the division-by-zero exception)
        without assistance from
        the \system software-level hypervisor.
        However,
        \system prevents a model core
        from directly interacting with GPUs or IO devices
        via port-mapped or memory-mapped IO.
        Instead,
        to issue an IO request,
        a model core writes the request
        a special IO DRAM region
        shared by the model and \system,
        and then raises an interrupt
        on a hypervisor core;
        the hypervisor core
        examines the request and,
        if the request is reasonable,
        interacts directly with the relevant IO device,
        placing the result in
        shared IO DRAM
        and then raising an interrupt
        on the initiating model core.
        To stop a model core
        from live-locking a hypervisor core
        with a flood of spurious interrupts,
        the LAPIC 
        chip
        of a hypervisor core
        throttles incoming requests,
        akin to the interrupt filter
        for an iPhone secure enclave processor~\cite{sepBlackhat2016}.
  \item Traditional virtualization-aware CPUs
        implement a special execution mode
        for 
        hypervisors.\footnote{For example,
        this mode is called ``root mode'' on Intel VT-x chips~\cite{uhlig2005},
        and ``hypervisor-extended supervisor mode''
        on RISC-V chips~\cite{risc-v-priv}.}
        Hypervisor modes enable
        same-core trap-and-emulate monitoring
        of how untrusted code
        uses sensitive instructions~\cite{popekGoldberg}.
        However,
        such modes are unnecessary in \system because
        a model core
        has no sensitive instructions
        to trap-and-emulate in the classic sense:
        hypervisor DRAM is physically inaccessible,
        the model core ISA has no instructions
        for accessing hypervisor core state,
        and IO devices can only be accessed
        indirectly via \system's port API
        (\S\ref{sec:hypervisor-sw}).
\end{itemize}
As explained in Section~\ref{sec:hypervisor-phys},
a hypervisor core
connects to a \system control terminal
via dedicated buses that are
inaccessible to model cores.

\system hardware
must support remote attestation~\cite{bootstrappingTrust}.
Before a model is loaded
onto a purported \system system
via the control terminal,
the terminal will verify that
the model is being sent
to valid \system silicon that runs
a valid \system software-level hypervisor.
All \system silicon
uses tamper-resistant and
tamper-evident technologies~\cite{impedanceVerif,backMon,immler2018,waksman2010}
to detect any sandbox circumventions
via model-launched social-engineering attacks
against hardware technicians;
hardware integrity should be
checked periodically by humans
as required by regulations (\S\ref{sec:hypervisor-policy}).
Note that checking for hardware integrity
not only requires inspection
of preexisting hardware,
but also verification that
no \textit{new} hardware has been added
via model-initiated social engineering,
e.g., to enable runaway model self-improvement.
\subsection{Software Hypervisor}
\label{sec:hypervisor-sw}


\system's software-level hypervisor
can be simpler than a traditional one
because it does not have to support
arbitrary kinds of guest code
that run on the same processors
which run hypervisor code.
\system's software-level hypervisor
only needs to be able to
supervise models which
run on model cores and
target the \textbf{\system guest API}.
The guest API
directly arises from
the affordances provided by
the \system hardware from Section~\ref{sec:hypervisor-uarch}:
\begin{itemize}[leftmargin=*]
  \item Model code can use any instruction
        provided by the model core ISA
        (including instructions for
        handling locally-gen\hyp erated
        interrupts and exceptions).
        Since the software-level hypervisor
        plays no role in handling such
        interrupts and exceptions,
        a model may choose to structure its code
        by distinguishing between
        OS software and user software,
        or by employing a unikernel approach~\cite{unikernel},
        or by doing something else.
        The \system software-level hypervisor
        is agnostic to a model's internal organization.
  \item Model cores have no way
        to directly contact devices
        besides model DRAM and model cores,
        so all requests to other kinds of devices
        must be sent to the memory region
        shared with hypervisor cores;
        the hypervisor cores will place results
        in that shared region.
\end{itemize}
The guest API interacts with devices (including GPUs)
via a \textbf{port API}.
\system ports are conceptually similar
to Mach ports~\cite{mach}.
Each port is a capability
that is granted by the software-level hypervisor
and which enables a model core
to interact with
a specific instance of a specific device type.
Each port maps to an address
in the DRAM region that
models share with the software-level hypervisor;
writing to that address
sends an interrupt to a hypervisor core,
with a model placing
additional information about the request
at a well-known place in the shared DRAM.
For example,
a port associated with a network device
might place a ring buffer~\cite{damn,rIOMMU}
in shared memory.
For security reasons,
\system explicitly disallows
models from directly engaging
with hardware via techniques
like SR-IOV~\cite{tu2015}---\system must
be able to synchronously monitor
all model/device interactions.
This ability,
together with the ability to
inspect model-core DRAM and registers,
allows \system to log
a model's inputs, outputs, and
intermediate states
for subsequent auditing
by the misbehavior detector.

Although \system treats the misbehavior detector
as a black box,
the affordances that \system provides
to hypervisor cores
are compatible with various ways of
identifying and fixing problematic model behavior.
For example:
\begin{itemize}[leftmargin=*]
  \item LLMs generate a response
        by using input tokens to
        traverse a path through the weights
        in a complex graph.
        Activation steering~\cite{soberLookAtSteering,representationEngineering,steeringLlama2}
        examines the weights
        that are triggered by each query,
        transforming a potentially dangerous model output
        into a less harmful one
        via on-the-fly substitution
        of the weights that are visited
        during the forward activation pass.
        Alternatively,
        a circuit-breaking approach~\cite{circuitBreaking}
        would disrupt a forward pass
        that visits problematic areas of the weight graph,
        preventing the model from
        generating any response at all.
        \system allows hypervisor cores to
        (1) introspect on each step of the forward pass,
        and (2) alter a model's intermediate state
        in arbitrary ways;
        \system does this while isolating hypervisor state
        that a misbehaving model
        might want to corrupt.
  \item In contrast to activation steering
        and circuit breaking
        (both of which require visibility into
        internal model state),
        input shielding~\cite{azurePromptShields,kyc} and
        output sanitization~\cite{phute2024,feuer2025}
        only examine a model's interactions
        with the outside world.
        Input shielding looks for suspicious prompts
        that might nudge a model towards misbehavior,
        whereas output sanitization
        removes problematic content from model responses.
        \system enables both approaches
        because \system restricts a model to using
        \system-controlled ports,
        such that all data exchanged via those ports
        is readable and writeable by hypervisor cores.
\end{itemize}
The best misbehavior detection/mitigation approaches
are model-specific,
and are dictated by the specific kinds of interactions
that the model has with external entities.

As described in Section~\ref{sec:hypervisor-uarch},
\system's software-level hypervisor
does not have to implement mechanisms
for virtualizing interrupts and exceptions~\cite{dall2018,vIC,tu2015}
on model cores.
\system's hypervisor also
need not implement
complex schedulers~\cite{flexcore,cypress}
or resource allocation schemes~\cite{gulati2010,bhargava2008}
to determine how to share model-core physical resources
among guests.
Once the hypervisor
has granted model cores and model DRAM
to a model,
the model is free to manage
the registers and memory
accessible to the model cores
in whatever way the model chooses;\footnote{Subject to
constraints imposed
when \system locks down a model core's MMU,
as described in Section~\ref{sec:hypervisor-uarch}.}
the model is responsible for
communicating with IO devices via
the \system port API,
so the hypervisor need not perform
device virtualization on model cores.
Keeping the hypervisor simple
helps to minimize
the hypervisor's threat surface.

To further minimize the threat surface,
\system's soft\hyp ware-level hypervisor
should be formally verified for correctness.
Projects like
sel4~\cite{sel4Whitepaper,sel4SOSP}
and mC2~\cite{certiKOS}
have demonstrated how hypervisors
can be proven free of
buffer overflows,
memory leaks,
non-terminating functions,
and other forms of misbehavior.
Writing formally verified software
is notoriously labor-intensive,
but the time investment
is worthwhile
to prevent existential AI risks.
If,
for whatever reason,
the hypervisor fails a software-level runtime assertion
or triggers an unexpected machine-check exception,
the hypervisor forcibly reboots
into offline isolation mode
(\S\ref{sec:hypervisor-phys}).

A \system hypervisor always uses
encrypted, authenticated network protocols
like TLS~\cite{tlsRFC} when communicating with
network hosts.
When a \system hypervisor
establishes such a connection,
the hypervisor explicitly announces itself
as being a \system hypervisor.
For example,
in the context of TLS,
the hypervisor's X.509 certificate~\cite{x509RFC},
issued and signed by an AI regulator (\S\ref{sec:hypervisor-policy}),
has an extension field indicating
that the certificate holder is a \system hypervisor;
during the TLS handshake,
the hypervisor will share the certificate
with the remote endpoint.
This self-identification as a \system hypervisor
allows the remote endpoint
to understand that its peer
is connected to a potentially dangerous AI model,
and thus should be treated
with a default level of suspicion.
Self-identification is particularly important
to prevent runaway model improvement
in which several models
communicate with each other
to collectively optimize themselves;
a \system hypervisor will refuse connection attempts
from other \system hypervisors.


\subsection{Physical Hypervisor}
\label{sec:hypervisor-phys}


\system has six isolation levels
for a sandboxed model:
\begin{enumerate}[leftmargin=*]
  \item {\bf Standard:} The model can receive any inputs
        and send any outputs, subject to the restrictions
        described in previous sections.
  \item {\bf Probation:} \system restricts
        model input and/or outputs,
        and/or performs additional logging,
        as directed by the control terminal.
  \item {\bf Severed:} \system prevents the model
        from using any ports,
        but leaves model cores powered-up
        so that hypervisor cores
        can examine model DRAM and registers,
        or perform higher-level interactions
        with the model via simulated IOs
        or direct manipulation of model state.
  \item {\bf Offline:} All model cores and hypervisor cores
        are completely powered down,
        and all support cables
        (e.g., network wires, power lines)
        are physically (but reversibly)
        disconnected from the external world.
  \item {\bf Decapitation:} Like offline isolation,
        except that support cables
        are physically damaged,
        such that they must be manually replaced
        to bring the model online again.
  \item {\bf Immolation:} All cores,
        support cables, HVAC systems,
        and other physical infrastructure
        are destroyed via fire, flooding,
        electromagnentic pulses,
        or other physical means.
\end{enumerate}
\system's physical hypervisor
consists of the \textbf{control console}
(an administrator machine connected to hypervisor cores
via dedicated buses),
the \textbf{kill switches}
(which implement actions
dictated by isolation levels 4, 5, or 6),
and the \textbf{misbehavior detector}
(which is part of the software-level hypervisor).
The control console
loads the software-level hypervisor
on hypervisor cores
and then tells it
which model to load on the model cores.
If the misbehavior detector
sends an alarm to the control console
(e.g., because of a machine check exception
on a hypervisor core,
or because of anomalous memory state
found on a model core),
the control console
orchestrates the transition to a new isolation model.
For example,
during a transition to probation isolation,
the control console might tell
the software-level hypervisor
how a model's port access should be restricted.
As another example,
during a transition to offline isolation,
the control console
orchestrates kill-switch activity
involving network cables, power lines, and so on.
Note that the software-level hypervisor
cannot independently transition the system
to a less-restrictive level,
only to a higher one.

The control console
can directly relax or restrict
the isolation level,
independent of a request from
the software-level hypervisor.
\system requires the control console
to have seven administrators.
Transitioning the deployment
to a less restrictive isolation mode
requires at least five-of-seven administrator consensus
via a hardware security module~\cite{thalesHSM};
transitioning to a more restrictive isolation level
only requires agreement
from at least three administrators.
These voting policies
create a bias towards safety,
and robustness against a malicious model
that has used social engineering
to corrupt a subset of \system administrators.

Hypervisor cores and the control console
exchange periodic heartbeats.
If a hypervisor core
fails to receive a heartbeat from the control console
(or vice versa),
\system transitions to offline isolation.


\subsection{Policy Hypervisor}
\label{sec:hypervisor-policy}
 

Governments have already begun to pass
laws which define
the types of models that
pose systemic risk.
For example,
in the EU AI Act,
such models have a high likelihood
``of harm to the health and safety
or the fundamental rights of persons,
taking into account both
the severity of the possible harm
and its probability of occurrence''~\cite{euAIact-recital52}.
Some 
specific threats
mentioned by the law are
nuclear, chemical, and biological harms,
as well as disinformation and
the automated discovery of cybersecurity vulnerabilities~\cite{euAIact-paragraph110}. 
The law
proposes several ways to
estimate a model's risk, e.g.,
by examining a model's parameter count
and training set size,
and by looking at the model's
level of autonomy~\cite{euAIact-annex13}.

We believe that
AI regulations should force
systemic-risk models
to run atop a \system-style hypervisor.
For example,
the EU AI Act requires
the operators of such models
to comply with reporting guidelines
like providing the European Commission
with technical documentation and model source code
when requested~\cite{euAIact-article92}.
Source code inspection
would provide evidence that
a model targets the \system guest API
(\S\ref{sec:hypervisor-sw}).
Regulators could also use
network-connected audit computers
to ask 
a live model
to attest that it uses
a \system hardware+software stack
(\S\ref{sec:hypervisor-uarch}).

Regulations should also mandate
in-person audits of the physical environment
for a \system hypervisor, e.g.,
to check the integrity of
tamper-resistant hardware enclosures,
and to verify that physical mechanisms
for implementing model decapitation and immolation (\S\ref{sec:hypervisor-phys})
are being properly maintained.
These audits
would draw inspiration from similar audits
used to verify
adherence to nuclear weapons treaties~\cite{start}
or standards for handling
the cryptographic material
used by HTTPS certificate authorities~\cite{mozilla,google}.

From the perspective of a model owner,
\system increases the cost of operating a model.
However,
regulators can incentivize the use of \system
(rather than just penalize its lack of use)
via ``safe harbor'' clauses~\cite{safeHarborDef} in AI laws.
These clauses
reduce a company's legal liability
if a company adhered to best practices
but nonetheless generated harm.

\section{Related Work}
\label{sec:related}

In the 1966 paper
``Speculations Concerning the First
Ultraintelligent Machine''~\cite{good1966},
Good introduced the term
``intelligence explosion'' to describe
the gains that would result
from a model which
surpassed the reasoning capacity
of a human.
Subsequent work
by philosophers and AI researchers
has considered whether a powerful model
could be placed inside
an ``AI box'' which limits
what the model could learn and do;
for an overview of this prior work,
see Yampolskiy's paper
on the AI confinement problem~\cite{leakproofingSingularity}.
To the best of our knowledge,
our own paper is the first
to detail what AI confinement
would mean at a fine-grained \textit{systems} level.

A prerequisite
for runaway model self-improvement
is the model being able to
introspect and update
model code and model weights---either its own,
or those of another model~\cite{turchin2018}.
However,
we are unaware of prior work
that has proposed fine-grained systems interventions
to prevent inappropriate model introspection
of model state.
For example,
Nevo et al. analyzed
various technical and social engineering mechanisms
by which model weights can be stolen~\cite{nevo2024}.
Nevo et al. then defined five security levels
for a model execution environment,
with higher levels imposing
increasingly strict operational requirements
on the execution environment.
For example,
at security level two and above,
model weights are stored exclusively on servers,
with no copies residing on personal devices;
at security levels four and five,
model inference executes within
confidential computing environments
akin to Graviton GPU enclaves~\cite{graviton}.
Nevo et al. did not provide
detailed technical proposals
for how to implement the different security levels,
whereas we explore concrete defensive mechanisms
like preventing model cores
from reading, modifying, and creating
executable pages or weight-containing pages.
Nevo et al. also
focused on model weight theft
as opposed to the more general problem
of sandboxing a malicious AI.

\vspace{-4mm}

\section{Final Thoughts}
At the time of this paper's writing,
AGI models do not exist.
As a result,
this paper's discussion about
AGI hypervisors
must necessarily suffer from
an evidence dilemma~\cite{bengio}---we
can only talk about AGI-induced harms
(and possible defenses against those harms)
in a speculative manner,
because the harm-inducing agents
do not currently exist.
However,
modern history is replete with examples
of societal challenges that
were identified in their nascent stages
and would have benefited from earlier,
more aggressive action.
For instance,
early predictions of global warning
were unfortunately prescient~\cite{theConversationClimateChange}
but did not capture the attention
of the public until decades later,
in some cases because energy companies 
suppressed the research findings of
their own scientists~\cite{harvardGazetteClimateChange};
climate scientists now debate
how quickly the ``point of no return''
will occur,
after which significant climate-related harms
will be unavoidable~\cite{aengenheyster2018}.
Social media technology
is another relevant example.
Concerns about
the technology's destabilizing impacts
(e.g., on mental health and
the dissemination of accurate information)
were known but deprioritized
by social media companies
and many of their computer scientists~\cite{facebookFiles}.

Technologists should not repeat these mistakes.
The systems community
(and computer scientists more generally)
should take an aggressively proactive approach
towards AI safety.
Policies for monitoring and restricting
potentially dangerous models require
systems-level enforcement mechanisms;
\system proposes one research direction
for such mechanisms.


{\footnotesize \bibliographystyle{acm}
\bibliography{refs}}

\begin{thebibliography}{10}

\bibitem{mach}
{\sc Accetta, M., Baron, R., Bolosky, W., Golub, D., Rashid, R., Tevanian, A., and Young, M.}
\newblock {Mach: A New Kernel Foundation For UNIX Development}.
\newblock In {\em Proceedings of the Summer USENIX Conference\/} (1986), pp.~93--112.

\bibitem{aengenheyster2018}
{\sc Aengenheyster, M., Feng, Q.~Y., van~der Ploeg, F., and Dijkstra, H.~A.}
\newblock The point of no return for climate action: effects of climate uncertainty and risk tolerance.
\newblock {\em Earth System Dynamics 9}, 3 (2018), 1085--1095.

\bibitem{vIC}
{\sc Ahmad, I., Gulati, A., and Mashtizadeh, A.}
\newblock {vIC: Interrupt Coalescing for Virtual Machine Storage Device IO}.
\newblock In {\em Proceedings of USENIX ATC\/} (June 2011), pp.~45--58.

\bibitem{kipScip}
{\sc {Apple}}.
\newblock {Apple Platform Security: Operating system integrity}.
\newblock \url{https://support.apple.com/guide/security/operating-system-integrity-sec8b776536b/web}, 2025.

\bibitem{bhargava2008}
{\sc Bhargava, R., Serebrin, B., Spadini, F., and Manne, S.}
\newblock {Accelerating Two-dimensional Page Walks for Virtualized Systems}.
\newblock In {\em Proceedings of ASPLOS\/} (March 2008), pp.~26--35.

\bibitem{x509RFC}
{\sc Boeyen, S., Santesson, S., Polk, T., Housley, R., Farrell, S., and Cooper, D.}
\newblock {Internet X.509 Public Key Infrastructure Certificate and Certificate Revocation List (CRL) Profile}.
\newblock RFC 5280, May 2008.

\bibitem{soberLookAtSteering}
{\sc Braun, J., Krasheninnikov, D., Anwar, U., Kirk, R., Tan, D., and Krueger, D.~S.}
\newblock {A Sober Look at Steering Vectors for LLMs}.
\newblock {LessWrong}, November 23, 2024.

\bibitem{safeHarborDef}
{\sc {Cornell Law School Legal Information Institute}}.
\newblock {Safe harbor}.
\newblock Legal Dictionary. \url{https://www.law.cornell.edu/wex/safe_harbor}, June, 2025.

\bibitem{clipper}
{\sc Crankshaw, D., Wang, X., Zhou, G., Franklin, M.~J., Gonzalez, J.~E., and Stoica, I.}
\newblock {Clipper: A Low-latency Online Prediction Serving System}.
\newblock In {\em Proceedings of NSDI\/} (March 2017), pp.~613--627.

\bibitem{dall2018}
{\sc Dall, C., Li, S.-W., Lim, J.~T., and Nieh, J.}
\newblock {ARM Virtualization: Performance and Architectural Implications}.
\newblock {\em SIGOPS Operating Systems Review 52}, 1 (August 2018), 45--56.

\bibitem{euAIact-annex13}
{\sc {EU Parliament}}.
\newblock {Artificial Intelligence Act (Regulation 2025/1689) - Annex XIII}, June 13, 2024.
\newblock \url{https://eur-lex.europa.eu/eli/reg/2024/1689}.

\bibitem{euAIact-article92}
{\sc {EU Parliament}}.
\newblock {Artificial Intelligence Act (Regulation 2025/1689) - Article 92}, June 13, 2024.
\newblock \url{https://eur-lex.europa.eu/eli/reg/2024/1689}.

\bibitem{euAIact-paragraph110}
{\sc {EU Parliament}}.
\newblock {Artificial Intelligence Act (Regulation 2025/1689) - Preamble, Paragraph 110}, June 13, 2024.
\newblock \url{https://eur-lex.europa.eu/eli/reg/2024/1689}.

\bibitem{euAIact-recital52}
{\sc {EU Parliament}}.
\newblock {Artificial Intelligence Act (Regulation 2025/1689) - Recital 52}, June 13, 2024.
\newblock \url{https://artificialintelligenceact.eu/recital/52/}.

\bibitem{cypress}
{\sc Fedorova, A., Kumar, V., Kazempour, V., Ray, S., and Alagheband, P.}
\newblock {Cypress: A Scheduling Infrastructure for a Many-Core Hypervisor}.
\newblock In {\em Proceedings of the Workshop on Managed Multi-Core Systems (MMCS)\/} (June 2008).

\bibitem{feuer2025}
{\sc Feuer, B., Goldblum, M., Datta, T., Nambiar, S., Besaleli, R., Dooley, S., Cembalest, M., and Dickerson, J.~P.}
\newblock {Style Outweighs Substance: Failure Modes of LLM Judges in Alignment Benchmarking}, January 27, 2025.
\newblock arXiv:2409.15268.

\bibitem{cachedAttention}
{\sc Gao, B., He, Z., Sharma, P., Kang, Q., , Jevdjic, D., Junbo~Deng, X.~Y., Yu, Z., and Zuo, P.}
\newblock {Cost-Efficient Large Language Model Serving for Multi-turn Conversations with CachedAttention}.
\newblock In {\em Proceedings of USENIX ATC\/} (July 2024), pp.~111--126.

\bibitem{ragSurvery2024}
{\sc Gao, Y., Xiong, Y., Gao, X., Jia, K., Pan, J., Bi, Y., Dai, Y., Sun, J., Wang, M., and Wang, H.}
\newblock {Retrieval-Augmented Generation for Large Language Models: A Survey}, March 27, 2024.
\newblock arXiv:2312.10997.

\bibitem{good1966}
{\sc Good, I.~J.}
\newblock {Speculations Concerning the First Ultraintelligent Machine}.
\newblock In {\em Advances in Computers}, F.~L. Alt and M.~Rubinoff, Eds., vol.~6. Elsevier, 1966, pp.~31--88.

\bibitem{google}
{\sc {Google Chrome}}.
\newblock {Chrome Root Program Policy, Version 1.5}.
\newblock \url{https://www.chromium.org/Home/chromium-security/root-ca-policy/}, January 2024.

\bibitem{greenblatt2024}
{\sc Greenblatt, R., Denison, C., Wright, B., Roger, F., MacDiarmid, M., Marks, S., Treutlein, J., Belonax, T., Chen, J., Duvenaud, D., Khan, A., Michael, J., Mindermann, S., Perez, E., Petrini, L., Uesato, J., Kaplan, J., Shlegeris, B., Bowman, S.~R., and Hubinger, E.}
\newblock Alignment faking in large language models, December 20, 2024.
\newblock arXiv:2412.14093.

\bibitem{kyc}
{\sc Greenblatt, R., and Shlegeris, B.}
\newblock {Managing Catastrophic Misuse Without Robust AIs}.
\newblock AI Alignment Forum. \url{https://www.alignmentforum.org/posts/KENtuXySHJgxsH2Qk/managing-catastrophic-misuse-without-robust-ais}, January 16, 2024.

\bibitem{certiKOS}
{\sc Gu, R., Shao, Z., Chen, H., Wu, X., Kim, J., Sj\"{o}berg, V., and Costanzo, D.}
\newblock {CertiKOS: An Extensible Architecture for Building Certified Concurrent OS Kernels}.
\newblock In {\em Proceedings of OSDI\/} (November 2016), pp.~653--669.

\bibitem{clockwork}
{\sc Gujarati, A., Karimi, R., Alzayat, S., Hao, W., Kaufmann, A., Vigfusson, Y., and Mace, J.}
\newblock {Serving DNNs Like Clockwork: Performance Predictability from the Bottom Up}.
\newblock In {\em Proceedings of OSDI\/} (November 2020), pp.~443--462.

\bibitem{gulati2010}
{\sc Gulati, A., Merchant, A., and Varman, P.~J.}
\newblock {mClock: Handling Throughput Variability for Hypervisor IO Scheduling}.
\newblock In {\em Proceedings of OSDI\/} (October 2010), pp.~437--450.

\bibitem{sel4Whitepaper}
{\sc Heiser, G.}
\newblock {The seL4 Microkernel: An Introduction}.
\newblock Revision 1.3. \url{https://beta.sel4.systems/About/seL4-whitepaper.pdf}, May 7, 2024.

\bibitem{bloom}
{\sc {Hugging Face}}.
\newblock {Introducing The World’s Largest Open Multilingual Language Model: BLOOM}.
\newblock \url{https://bigscience.huggingface.co/blog/bloom}, July 12, 2024.

\bibitem{immler2018}
{\sc Immler, V., Obermaier, J., Ng, K.~K., Ke, F.~X., Lee, J., Lim, Y.~P., Oh, W.~K., Wee, K.~H., and Sigl, G.}
\newblock {Secure Physical Enclosures from Covers with Tamper-Resistance}.
\newblock {\em Transactions on Cryptographic Hardware and Embedded Systems 2019\/} (November 2018).

\bibitem{ragCache}
{\sc Jin, C., Zhang, Z., Jiang, X., Liu, F., Liu, X., Liu, X., and Jin, X.}
\newblock Ragcache: Efficient knowledge caching for retrieval-augmented generation, April 25, 2024.
\newblock arXiv:2404.12457.

\bibitem{sel4SOSP}
{\sc Klein, G., Elphinstone, K., Heiser, G., Andronick, J., Cock, D., Derrin, P., Elkaduwe, D., Engelhardt, K., Kolanski, R., Norrish, M., Sewell, T., Tuch, H., and Winwood, S.}
\newblock {seL4: Formal Verification of an OS Kernel}.
\newblock In {\em Proceedings of SOSP\/} (2009), pp.~207--220.

\bibitem{spectre}
{\sc Kocher, P., Horn, J., Fogh, A., Genkin, D., Gruss, D., Haas, W., Hamburg, M., Lipp, M., Mangard, S., Prescher, T., Schwarz, M., and Yarom, Y.}
\newblock {Spectre Attacks: Exploiting Speculative Execution}.
\newblock In {\em Proceedings of the IEEE Symposium on Security and Privacy\/} (May 2019), pp.~1--19.

\bibitem{vLLM}
{\sc Kwon, W., Li, Z., Zhuang, S., Sheng, Y., Zheng, L., Yu, C.~H., Gonzalez, J., Zhang, H., and Stoica, I.}
\newblock Efficient memory management for large language model serving with pagedattention.
\newblock In {\em Proceedings of SOSP\/} (October 2023), pp.~611--626.

\bibitem{lin-greenberg-2020}
{\sc Lin-Greenberg, E.}
\newblock {Allies and Artificial Intelligence: Obstacles to Operations and Decision-Making}.
\newblock {\em Texas National Security Review 3}, 2 (2020), 56--76.

\bibitem{unikernel}
{\sc Madhavapeddy, A., Mortier, R., Rotsos, C., Scott, D., Singh, B., Gazagnaire, T., Smith, S., Hand, S., and Crowcroft, J.}
\newblock {Unikernels: Library Operating Systems for the Cloud}.
\newblock In {\em Proceedings of ASPLOS\/} (March 2013), pp.~461--472.

\bibitem{rIOMMU}
{\sc Malka, M., Amit, N., Ben-Yehuda, M., and Tsafrir, D.}
\newblock {rIOMMU: Efficient IOMMU for I/O Devices that Employ Ring Buffers}.
\newblock In {\em Proceedings of ASPLOS\/} (March 2015), pp.~355--368.

\bibitem{damn}
{\sc Markuze, A., Smolyar, I., Morrison, A., and Tsafrir, D.}
\newblock {DAMN: Overhead-Free IOMMU Protection for Networking}.
\newblock In {\em Proceedings of ASPLOS\/} (March 2018), pp.~301--315.

\bibitem{harvardGazetteClimateChange}
{\sc McCarthy, A.}
\newblock {Exxon disputed climate findings for years. Its scientists knew better.}, January 12, 2023.
\newblock \url{https://news.harvard.edu/gazette/story/2023/01/harvard-led-analysis-finds-exxonmobil-internal-research-accurately-predicted-climate-change/}.

\bibitem{llama3-1}
{\sc {Meta}}.
\newblock {Introducing Llama 3.1: Our most capable models to date}.
\newblock \url{https://ai.meta.com/blog/meta-llama-3-1/}, July 23, 2024.

\bibitem{flexcore}
{\sc Miao, T., and Chen, H.}
\newblock {FlexCore: Dynamic Virtual Machine Scheduling using VCPU Ballooning}.
\newblock {\em Tsinghua Science and Technology 20}, 1 (2015), 7--16.

\bibitem{azurePromptShields}
{\sc {Microsoft}}.
\newblock {Prompt Shields}.
\newblock \url{https://learn.microsoft.com/en-us/azure/ai-services/content-safety/concepts/jailbreak-detection}, October 17, 2024.

\bibitem{min2022}
{\sc Min, B.~H., and Borch, C.}
\newblock {Systemic failures and organizational risk management in algorithmic trading: Normal accidents and high reliability in financial markets}.
\newblock {\em Social Studies of Science 52}, 2 (2022), 277--302.

\bibitem{rand-military-ai}
{\sc Morgan, F.~E., Boudreaux, B., Lohn, A.~J., Ashby, M., Curriden, C., Klima, K., and Grossman, D.}
\newblock {\em Military Applications of Artificial Intelligence: Ethical Concerns in an Uncertain World}.
\newblock RAND Corporation, Santa Monica, CA, 2020.

\bibitem{impedanceVerif}
{\sc Mosavirik, T., Schaumont, P., and Tajik, S.}
\newblock {ImpedanceVerif: On-Chip Impedance Sensing for System-Level Tampering Detection}.
\newblock {\em IACR Transactions on Cryptographic Hardware and Embedded Systems 2023}, 1 (November 2022), 301--325.

\bibitem{backMon}
{\sc Mosavirik, T., and Tajik, S.}
\newblock {BackMon: IC Backside Tamper Detection using On-Chip Impedance Monitoring}.
\newblock In {\em Proceedings of the Workshop on Attacks and Solutions in Hardware Security\/} (October 2024), pp.~68--77.

\bibitem{mozilla}
{\sc {Mozilla}}.
\newblock {Mozilla's CA Certificate Program}.
\newblock \url{https://www.google.com/url?q=https://www.mozilla.org/en-US/about/governance/policies/security-group/certs/policy/&source=gmail&ust=1736239187085000&usg=AOvVaw2lv3d3uS6qLJwSmkjqL4CE }, December 2024.

\bibitem{nevo2024}
{\sc Nevo, S., Lahav, D., Karpur, A., Bar-on, Y., Brardley, H.~A., and Alstott, J.}
\newblock {Securing AI Model Weights: Preventing Theft and Misuse of Frontier Models}, May 30, 2024.

\bibitem{theConversationClimateChange}
{\sc Nicholls, N.}
\newblock {40 years ago, scientists predicted climate change. And hey, they were right.}, July 28, 2019.
\newblock \url{https://theconversation.com/40-years-ago-scientists-predicted-climate-change-and-hey-they-were-right-120502}.

\bibitem{marconi}
{\sc Pan, R., Wang, Z., Jia, Z., Karakus, C., Zancato, L., Dao, T., Wang, Y., and Netravali, R.}
\newblock {Marconi: Prefix Caching for the Era of Hybrid LLMs}, December 4, 2024.
\newblock arXiv:2411.19379.

\bibitem{steeringLlama2}
{\sc Panickssery, N., Gabrieli, N., Schulz, J., Tong, M., Hubinger, E., and Turner, A.~M.}
\newblock {Steering Llama 2 via Contrastive Activation Addition}, July 5, 2024.
\newblock arXiv:2312.06681.

\bibitem{bootstrappingTrust}
{\sc Parno, B., McCune, J.~M., and Perrig, A.}
\newblock {\em {Bootstrapping Trust in Modern Computers}}, 1st~ed.
\newblock Springer, 2011.

\bibitem{phute2024}
{\sc Phute, M., Helbling, A., Hull, M., Peng, S., Szyller, S., Cornelius, C., and Chau, D.~H.}
\newblock {LLM Self Defense: By Self Examination, LLMs Know They Are Being Tricked}, May 2, 2024.
\newblock arXiv:2308.07308.

\bibitem{popekGoldberg}
{\sc Popek, G.~J., and Goldberg, R.~P.}
\newblock {Formal requirements for virtualizable third generation architectures}.
\newblock {\em Commununications of the ACM 17}, 7 (July 1974), 412--421.

\bibitem{ragServe}
{\sc Ray, S., Pan, R., Gu, Z., Du, K., Ananthanarayanan, G., Netravali, R., and Jiang, J.}
\newblock {RAGServe: Fast Quality-Aware RAG Systems with Configuration Adaptation}, December 13, 2024.
\newblock arXiv:2412.10543.

\bibitem{tlsRFC}
{\sc Rescorla, E.}
\newblock {The Transport Layer Security (TLS) Protocol Version 1.3}.
\newblock RFC 8446, August 2018.

\bibitem{risc-v-priv}
{\sc {RISC-V}}.
\newblock {The RISC-V Instruction Set Manual: Volume II (Privileged Architecture). Version 20241101}.
\newblock \url{https://github.com/riscv/riscv-isa-manual/releases/download/riscv-isa-release-7c5adda-2025-01-02/riscv-privileged.pdf}, November 2024.

\bibitem{salinas2024}
{\sc Salinas, A., and Morstatter, F.}
\newblock {The Butterfly Effect of Altering Prompts: How Small Changes and Jailbreaks Affect Large Language Model Performance}, January 9, 2024.
\newblock arXiv:2401.03729v2.

\bibitem{vicarte2021}
{\sc Sanchez~Vicarte, J.~R., Shome, P., Nayak, N., Trippel, C., Morrison, A., Kohlbrenner, D., and Fletcher, C.~W.}
\newblock {Opening Pandora’s Box: A Systematic Study of New Ways Microarchitecture Can Leak Private Data}.
\newblock In {\em Proceedings of ISCA\/} (2021), pp.~347--360.

\bibitem{chatgpt-params}
{\sc Schreiner, M.}
\newblock {GPT-4 architecture, datasets, costs and more leaked}.
\newblock Decoder. \url{https://the-decoder.com/gpt-4-architecture-datasets-costs-and-more-leaked/}, July 11, 2023.

\bibitem{nexus}
{\sc Shen, H., Chen, L., Jin, Y., Zhao, L., Kong, B., Philipose, M., Krishnamurthy, A., and Sundaram, R.}
\newblock {Nexus: A GPU Cluster Engine for Accelerating DNN-based Video Analysis}.
\newblock In {\em Proceedings of SOSP\/} (2019), pp.~322--337.

\bibitem{singh2024}
{\sc Singh, C., Inala, J.~P., Galley, M., Caruana, R., and Gao, J.}
\newblock {Rethinking Interpretability in the Era of Large Language Models}, January 30, 2024.
\newblock arXiv:2402.01761v.

\bibitem{sa2023}
{\sc Sá, B., Valente, L., Martins, J., Rossi, D., Benini, L., and Pinto, S.}
\newblock {CVA6 RISC-V Virtualization: Architecture, Microarchitecture, and Design Space Exploration}.
\newblock {\em {IEEE Transactions on Very Large Scale Integration (VLSI) Systems} 31}, 11 (2023), 1713--1726.

\bibitem{sepBlackhat2016}
{\sc {Tarjei Mandt and Mathew Solnik and David Wang}}.
\newblock {Demystifying the Secure Enclave Processor}.
\newblock BlackHat USA. \url{https://mista.nu/research/sep-paper.pdf}, 2015.

\bibitem{thalesHSM}
{\sc {Thales}}.
\newblock {Luna USB HSM Documentation: Multifactor Quorum Authentication}.
\newblock \url{https://thalesdocs.com/gphsm/luna/7/docs/usb/Content/admin_usb/hsm/multifactor/multifactor_auth.htm}, 2024.

\bibitem{das2024}
{\sc Tonmoy, S., Zaman, S., Jain, V., Rani, A., Rawte, V., Chadha, A., and Das, A.}
\newblock {A comprehensive survey of hallucination mitiga- tion techniques in large language models}, January 8, 2024.
\newblock arXiv:2401.01313.

\bibitem{tu2015}
{\sc Tu, C.-C., Ferdman, M., Lee, C.-t., and Chiueh, T.-c.}
\newblock {A Comprehensive Implementation and Evaluation of Direct Interrupt Delivery}.
\newblock In {\em Proceedings of VEE\/} (March 2015), pp.~1--15.

\bibitem{smt}
{\sc Tullsen, D., Eggers, S., and Levy, H.}
\newblock {Simultaneous multithreading: Maximizing On-chip Parallelism}.
\newblock In {\em Proceedings of ISCA\/} (June 1995), pp.~392--403.

\bibitem{turchin2018}
{\sc Turchin, A.}
\newblock {Levels of AI Self-Improvement}.
\newblock \url{https://www.lesswrong.com/posts/os7N7nJoezWKQnnuW/levels-of-ai-self-improvement}, April 29, 2018.

\bibitem{turpin2023}
{\sc Turpin, M., Michael, J., Perez, E., and Bowman, S.~R.}
\newblock {Language models don’t always say what they think: Unfaithful explanations in chain-of-thought prompting}, December 9, 2024.
\newblock arXiv:2305.04388.

\bibitem{uhlig2005}
{\sc Uhlig, R., Neiger, G., Rodgers, D., Santoni, A., Martins, F., Anderson, A., Bennett, S., Kagi, A., Leung, F., and Smith, L.}
\newblock {Intel virtualization technology}.
\newblock {\em {IEEE Computer} 38}, 5 (2005), 48--56.

\bibitem{bengio}
{\sc {UK AI Safety Institute}}.
\newblock {International AI Safety Report: The International Scientific Report on the Safety of Advanced AI}, February 18 2025.
\newblock \url{https://www.gov.uk/government/publications/international-ai-safety-report-2025}.

\bibitem{start}
{\sc {United States Government Accountability Office}}.
\newblock {NUCLEAR ARMS CONTROL: U.S. May Face Challenges in Verifying Future Treaty Goals}.
\newblock \url{https://www.gao.gov/assets/gao-23-105698.pdf}, September 2023.

\bibitem{graviton}
{\sc Volos, S., Vaswani, K., and Bruno, R.}
\newblock {Graviton: Trusted Execution Environments on GPUs }.
\newblock In {\em Proceedings of the USENIX Symposium on Operating Systems Design and Implementation\/} (October 2018), pp.~681--696.

\bibitem{waksman2010}
{\sc Waksman, A., and Sethumadhavan, S.}
\newblock {Tamper Evident Microprocessors}.
\newblock In {\em Proceedings of the IEEE Symposium on Security and Privacy\/} (May 2010), pp.~173--188.

\bibitem{facebookFiles}
{\sc {Wall Street Journal}}.
\newblock {The Facebook Files}, January 12, 2022.
\newblock \url{https://www.wsj.com/tech/the-facebook-files-11642035385}.

\bibitem{foreshadow-ng}
{\sc Weisse, O., Van~Bulck, J., Minkin, M., Genkin, D., Kasikci, B., Piessens, F., Silberstein, M., Strackx, R., Wenisch, T.~F., and Yarom, Y.}
\newblock {Foreshadow-NG}: Breaking the virtual memory abstraction with transient out-of-order execution.
\newblock {\em Technical report, revision 1.0\/} (August 14 2018).

\bibitem{wistoff21}
{\sc Wistoff, N., Schneider, M., Gürkaynak, F.~K., Benini, L., and Heiser, G.}
\newblock {Microarchitectural Timing Channels and their Prevention on an Open-Source 64-bit RISC-V Core}.
\newblock In {\em Proceedings of the Design, Automation, and Test in Europe Conference (DATE)\/} (February 2021), pp.~627--632.

\bibitem{wolf2024}
{\sc Wolf, Y., Wies, N., Avnery, O., Levine, Y., and Shashua, A.}
\newblock {Fundamental Limitations of Alignment in Large Language Models}, June 3, 2024.
\newblock arXiv:2304.11082.

\bibitem{xie2022}
{\sc Xie, Y., Wang, D., Chen, P.-Y., Xiong, J., Liu, S., and Koyejo, O.}
\newblock {A Word is Worth A Thousand Dollars: Adversarial Attack on Tweets Fools Stock Prediction}.
\newblock In {\em Proceedings of the 2022 Conference of the North American Chapter of the Association for Computational Linguistics: Human Language Technologies\/} (Seattle, WA, July 2022).

\bibitem{leakproofingSingularity}
{\sc Yampolskiy, R.}
\newblock {Leakproofing the Singularity: Artificial Intelligence Confinement Problem}.
\newblock {\em Journal of Consciousness Studies 19}, 1-2 (2012), 194--214.

\bibitem{durrett2022}
{\sc Ye, X., and Durrett, G.}
\newblock {The unreliability of explanations in few-shot prompting for textual reasoning}.
\newblock In {\em Proceedings of NeurIPS\/} (December 2022).

\bibitem{shepherd}
{\sc Zhang, H., Tang, Y., Khandelwal, A., and Stoica, I.}
\newblock {SHEPHERD: Serving DNNs in the Wild}.
\newblock In {\em Proceedings of NSDI\/} (April 2023), pp.~787--808.

\bibitem{representationEngineering}
{\sc Zou, A., Phan, L., Chen, S., Campbell, J., Guo, P., Ren, R., Pan, A., Yin, X., Mazeika, M., Dombrowski, A.-K., Goel, S., Li, N., Byun, M.~J., Wang, Z., Mallen, A., Basart, S., Koyejo, S., Song, D., Fredrikson, M., Kolter, J.~Z., and Hendrycks, D.}
\newblock {Representation Engineering: A Top-Down Approach to AI Transparency}, March 3, 2025.
\newblock arXiv:2310.01405.

\bibitem{circuitBreaking}
{\sc Zou, A., Phan, L., Wang, J., Duenas, D., Lin, M., Andriushchenko, M., Wang, R., Kolter, Z., Fredrikson, M., and Hendrycks, D.}
\newblock {Improving Alignment and Robustness with Circuit Breakers}, July 12, 2024.
\newblock arXiv:2406.04313.

\end{thebibliography}


\end{document}